\documentclass[aps,prd,amsmath,amssymb,amsfonts,reprint,nofootinbib]{revtex4-1}

\usepackage{graphicx}

\graphicspath{{./img/}}

\newcommand{\iu}{{i\mkern1mu}}
\newcommand{\inneri}[2]{\ensuremath{\,\left(\,{#1}\,|\,{#2}\,\right)\,}}
\newcommand{\innerii}[2]{\ensuremath{\,\left\langle\,{#1}\,|\,{#2}\,\right\rangle\,}}
\newcommand{\dd}{\ensuremath{\mathrm{d}}}
\newcommand{\intd}[4]{\ensuremath{\int_{#1}^{#2}{#3}\,\dd{#4}}}
\newcommand{\partialdiff}[2]{\ensuremath{\dfrac{\partial {#1}}{\partial {#2}}}}

\newcommand{\figref}[1]{\ref{fig:#1}}

\DeclareMathOperator\erf{erf}

\begin{document}

\title{Likelihood smoothing using gravitational wave surrogate models}

\author{Robert H. Cole}
\email[]{rhc26@ast.cam.ac.uk}
\author{Jonathan R. Gair}
\affiliation{Institute of Astronomy, Madingley Road, Cambridge, CB3 0HA, United Kingdom}

\date{December 16, 2014}

\begin{abstract}
Likelihood surfaces in the parameter space of gravitational wave signals can contain many secondary maxima, which can prevent search algorithms from finding the global peak and correctly mapping the distribution. Traditional schemes to mitigate this problem maintain the number of secondary maxima and thus retain the possibility that the global maximum will remain undiscovered. By contrast, the recently proposed technique of likelihood transform can modify the structure of the likelihood surface to reduce its complexity. We present a practical method to carry out a likelihood transform using a Gaussian smoothing kernel, utilising gravitational wave surrogate models to perform the smoothing operation analytically. We demonstrate the approach with Newtonian and post-Newtonian waveform models for an inspiralling circular compact binary.
\end{abstract}


\maketitle

\section{\label{sec:intro}Introduction}
The first direct detection of gravitational waves (GWs) is likely to occur soon. The ground-based interferometers LIGO~\cite{Abbott2009} and Virgo~\cite{Accadia2012} are currently undergoing upgrades to Advanced configurations which should start taking data in the next couple of years and, when they achieve their final design sensitivity, are expected to detect GWs from the inspiral and merger of stellar compact binaries at the rate of several events per year~\cite{Abadie2010}. There are ongoing efforts to detect a stochastic background of nanohertz GWs generated by merging supermassive black hole binaries using the accurate timing of arrays of millisecond pulsars and the first results could come within five years~\cite{Sesana2013}. Further in the future, the European Space Agency has selected ``The Gravitational Universe'' to be the science theme for the L3 science mission to launch in 2034, which aims to detect millihertz GWs using a space-based interferometer~\cite{Amaro-Seoane2012}. These distinct efforts to measure GWs span the wide frequency range of potential sources~\cite{Moore2014} and it is expected that GW observations from many different sources will eventually become routine~\cite{Harry2010}.

For many astrophysical systems, we can produce accurate models of the source and hence predict the waveform that would be observed on Earth. Given a bank of predicted waveform templates, a GW detection can be made using matched filtering: the comparison of observed data with every template in the bank. For this to be an effective technique, the templates need to be closely separated in parameter space; in high dimensional spaces, it is not possible to construct a template bank of sufficient density using reasonable computational resources. Rather, it is common to use a Markov chain Monte Carlo (MCMC) method to map the \emph{posterior} distribution, evaluating waveforms as required.

The posterior is the probability of a particular set of model parameters given some observed data. To achieve a mapping of the posterior surface with high enough resolution requires many evaluations of the \emph{likelihood} function, which is the probability of producing a particular data stream given some waveform model parameters. The Metropolis-Hastings algorithm provides a technique to explore the parameter space using a random walk, producing a chain of samples that will eventually converge to the posterior distribution. There is no guarantee about how long this convergence will take and the number of likelihood evaluations may still be large. Techniques to accelerate the convergence of MCMC routines are highly desirable.

Simulated annealing~\cite{Kirkpatrick1983} is one such method. The idea is to ``heat up'' the likelihood surface by replacing $\mathcal{L}$ with $\mathcal{L}_T=\mathcal{L}^{1/T}$ for some temperature $T > 1$, making it easier for chains to explore large regions of parameter space. Following some predetermined cooling schedule, the temperature is gradually reduced to $T=1$, where the chain begins to sample the true likelihood, starting from a point that is more likely to be near the global maximum. This approach is advantageous because it does not slow down the likelihood evaluation at each point, but the number of local maxima of $\mathcal{L}$ remains fixed. Within some fixed computational time, it is possible for a chain to remain close to a secondary and not find the global maximum.

Parallel tempering is related to simulated annealing in that it uses the modified likelihoods $\mathcal{L}_T$. Chains are run simultaneously on a ladder of different temperatures, with high $T$ chains exploring more of the parameter space. Swaps between the locations of adjacent chains are proposed, and in this way information about the global structure of the surface is propagated down to the $T=1$ chain, which is sampling the desired distribution.

Likelihood transform techniques~\cite{Wang2014} were recently suggested as an alternative and aim to accelerate MCMC convergence by modifying the likelihood surface in a more complicated way; specifically, we consider the case where $\mathcal{L}$ is convolved with a smoothing kernel $\mathcal{K}_\sigma$. This reduces the number of local maxima, but at the cost of an increased evaluation time at each point.

A separate approach to the problem is to speed up the individual likelihood evaluations by using reduced order methods~\cite{Canizares2013}. These accelerate the likelihood calculation by reducing the number of time or frequency samples at which the waveforms need to be evaluated by first finding a reduced spanning set for the waveform space. Similarly, the construction of surrogate models~\cite{Puerrer2014,Field2014} achieves acceleration of waveform computations by interpolating the waveform space. In both approaches, significant computational work is done offline to produce the interpolations, allowing for a quicker online run-time.

In this paper, we combine the principles of waveform interpolation and likelihood transformation via a smoothing convolution. A practical scheme for applying likelihood transform methods to GW data analysis with surrogate models is developed, allowing for accelerated convergence in MCMC searches without additional computational time. In section \ref{sec:GWs}, we introduce our notation for gravitational waveforms and outline the generation of surrogate models. The principle behind likelihood transform techniques is discussed in section \ref{sec:likelihood-transform}, along with the application of surrogate models to this problem. We then present some specific examples in section \ref{sec:examples}, before concluding in section \ref{sec:conclusions} with a discussion. 

\section{\label{sec:GWs}Gravitational Waves}
A gravitational waveform $h(t; \boldsymbol{\lambda})$, depending on some set of parameters $\boldsymbol{\lambda}$, has two independent components, describing two polarisations: plus $+$ and cross $\times$. A given GW detector is sensitive to a particular linear combination of the waveform polarisations
\begin{equation}
h_\alpha(t;\boldsymbol{\lambda}) = F^+_\alpha h_+(t; \boldsymbol{\lambda}) + F^\times_\alpha h_\times(t; \boldsymbol{\lambda}),
\end{equation}
where a subscript $\alpha$ denotes a specific detector and the response functions $F^A_\alpha$ depend on the relative orientations of the detector and the GW source. For the initial generation of ground-based detectors, these are essentially constant over the duration of a typical signal, but they may vary significantly over an observation for space-based interferometers as well as advanced ground-based detectors.

In data analysis, it is necessary to account for the noise present in GW detectors, which is assumed to be stationary and Gaussian. The natural overlap between waveforms is in the frequency domain
\begin{equation}
\label{eq:noise-overlap}
\innerii{h(\boldsymbol{\lambda}_1)}{h(\boldsymbol{\lambda}_2)} \equiv 4\sum_{\alpha}\intd{0}{\infty}{\frac{\tilde{h}^*_\alpha(f; \boldsymbol{\lambda}_1) \tilde{h}_\alpha(f; \boldsymbol{\lambda}_2)}{S_{n,\alpha}(f)}}{f},
\end{equation}
where we sum over different detectors, each with their own one-sided noise power spectral density $S_{n,\alpha}(f)$. Here $\tilde{h}_\alpha(f)$ denotes the Fourier transform of $h_\alpha(t)$, and we are using $\innerii{\cdot}{\cdot}$ to denote a noise-weighted overlap.

It is often convenient to write the waveform as a complex time series
\begin{equation}
h(t; \boldsymbol{\lambda}) = h_+(t; \boldsymbol{\lambda}) + \iu h_\times(t; \boldsymbol{\lambda}).
\end{equation}
The natural inner product on this complex waveform space is
\begin{equation}
\inneri{h(\boldsymbol{\lambda}_1)}{h(\boldsymbol{\lambda}_2)} \equiv \intd{-\infty}{\infty}{h^*(t; \boldsymbol{\lambda}_1) h(t; \boldsymbol{\lambda}_2)}{t},
\end{equation}
where in practice, the integral is of finite length, $T$, equal to the observation time. The corresponding real overlap between two waveforms is given by the real part of this inner product. The complex inner product and associated overlap make no reference to a particular detector and are therefore useful for constructing reduced spanning sets for waveform spaces~\cite{Field2014}. The two overlaps coincide if it is assumed that there are two right-angled detectors, at $45^\circ$ to one another, with independent white noise, $S_{n,1}=S_{n,2} =\mbox{const.}$, and that the source is optimally oriented, i.e., the principal polarisation axes of the source are aligned with the arms of the first detector. In this configuration the detector aligned with the principal axes is sensitive only to the plus polarisation, while the other detector is sensitive only to the cross polarisation and the overlap is proportional to
\begin{equation}
\sum_{A=+,\times}\intd{-\infty}{\infty}{h_A(t; \boldsymbol{\lambda}_1) h_A(t; \boldsymbol{\lambda}_2)}{t} = \Re[\inneri{h(\boldsymbol{\lambda}_1)}{h(\boldsymbol{\lambda}_2)}],
\end{equation}
where the sum is over polarisation states. We will assume this optimal configuration for all sources in subsequent calculations, as the likelihood transform technique is independent of our choice of detector.

\subsection{Data analysis}
Given some measured detector data $x_\alpha(t) = h_\alpha(t; \boldsymbol{\lambda}_*) + n_\alpha(t)$, composed of a GW signal $h$ with parameters $\boldsymbol{\lambda}_*$ and detector noise $n$, the signal-to-noise ratio (SNR) can be calculated
\begin{equation}
\rho(\boldsymbol{\lambda}) = \frac{\innerii{x}{h(\boldsymbol{\lambda})}}{\sqrt{\innerii{h(\boldsymbol{\lambda})}{h(\boldsymbol{\lambda})}}},
\end{equation}
which we expect to be strongly peaked at the true parameters $\boldsymbol{\lambda}_*$. It is useful to work with normalised templates, such that $\innerii{h(\boldsymbol{\lambda})}{h(\boldsymbol{\lambda})} = 1$. For noise-free data, $x_\alpha(t) = h_\alpha(t;\boldsymbol{\lambda}_*) $, and making the simplifying assumptions about the source orientation and noise properties described above, the SNR reduces to
\begin{equation}
\label{eq:SNR}
\rho(\boldsymbol{\lambda}; \boldsymbol{\lambda}_*) = \Re[\inneri{h(\boldsymbol{\lambda}_*)}{h(\boldsymbol{\lambda})}],
\end{equation}
which is linear in the model $h(\boldsymbol{\lambda})$.

The likelihood $\mathcal{L}(x | \boldsymbol{\lambda})$ is the probability that a particular data stream $x$ is observed, given the parameters $\boldsymbol{\lambda}$ of the signal present. Assuming stationary Gaussian noise, the likelihood is simply
\begin{equation}\label{eq:likelihood}
\mathcal{L}(x | \boldsymbol{\lambda}) \propto \exp\left[-\innerii{x-h(\boldsymbol{\lambda})}{x-h(\boldsymbol{\lambda})}/\,2\right],
\end{equation}
up to some normalising factor. The actual quantity of interest is the posterior $\mathcal{P}(\boldsymbol{\lambda}| x)$, which is the probability distribution of the parameters of the source $\boldsymbol{\lambda}$, given the observed data stream $x$. It is related to $\mathcal{L}$ via Bayes' theorem
\begin{equation}
\mathcal{P}(\boldsymbol{\lambda}| x) = \frac{\mathcal{L}(x|\boldsymbol{\lambda})\pi(\boldsymbol{\lambda})}{Z}, \qquad Z = \int {\rm d}\boldsymbol{\lambda}\,\, \mathcal{L}(x|\boldsymbol{\lambda})\pi(\boldsymbol{\lambda})
\end{equation}
where $\pi(\boldsymbol{\lambda})$ is the prior probability distribution for the parameters, which reflects our beliefs about the source parameters prior to the data being taken. The evidence, $Z$, normalises the posterior over the parameter space and can also be used for model selection.

The expectation value of the likelihood over different noise realisations can be expanded about the true parameters in the linear signal approximation to yield the Fisher Information Matrix (FIM)
\begin{equation}
\Gamma_{ij} = \innerii{h_{,i}}{h_{,j}},
\end{equation}
where $h_{,i} = \partial h/\partial\lambda_i$ denotes the partial derivative of the waveform model with respect to the parameters. In the case of uninformative (uniform) priors, the inverse of the FIM is the covariance of the posterior distribution, and so can be used to set a scale on its structure. The component $(\Gamma^{-1})_{ii}$ is a measure of the expected width of the marginalised posterior in $\lambda_i$ and hence is a measure of the expected uncertainty in the measurement of that parameter from the observed data.

The posterior encodes all of the information about the source that can be determined from an observation. In low-dimensional parameter spaces, the posterior can be evaluated on a fine grid in parameter space. For higher dimensional spaces, the required computational time is too large and alternative methods are typically adopted, such as Markov chain Monte Carlo methods. 

MCMC techniques aim to generate a chain of samples, $\boldsymbol{\lambda}_i$ in which, after a burn-in phase, the density of points is proportional to the posterior distribution. This is typically achieved using the Metropolis-Hastings algorithm. Given some current parameter value $\boldsymbol{\lambda}_i$, a new candidate parameter value $\boldsymbol{\lambda}'$ is chosen from a suitable proposal distribution $q(\boldsymbol{\lambda}'|\boldsymbol{\lambda}_i)$, for instance a Gaussian centred at $\boldsymbol{\lambda}_i$. The Metropolis-Hastings ratio 
\begin{equation}
\alpha = \frac{\mathcal{P}(\boldsymbol{\lambda}'| x) q(\boldsymbol{\lambda}_i|\boldsymbol{\lambda}')}{\mathcal{P}(\boldsymbol{\lambda}_{i}| x) q(\boldsymbol{\lambda}'|\boldsymbol{\lambda}_i)}
\end{equation}
is calculated and the new state is accepted with probability $\min(1,\alpha)$. Otherwise, the next state is set to be the current parameter values $\boldsymbol{\lambda}_{i+1}=\boldsymbol{\lambda}_i$. The starting point for the algorithm, $\boldsymbol{\lambda}_0$, can be chosen arbitrarily.

\subsection{\label{sec:surrogate}Reduced order methods and surrogate models}
Gravitational waveforms are routinely generated for arbitrary system parameters by numerically solving differential equations. The accuracy of such waveforms is not guaranteed (and in some cases, deliberately sacrificed to reduce the computational cost, for example kludge models of extreme-mass-ratio inspirals~\cite{Barack2004,Babak2007}); indeed, it is possible to get significant systematic biases in parameter estimation by using unfaithful waveform templates, but we shall assume in this analysis that the numerical templates can be calculated exactly. Waveform models and the associated likelihood, Eq.~(\ref{eq:likelihood}), can be expensive to evaluate and reduced order methods have been proposed as a way to speed up such likelihood evaluations. These rely on finding an approximation to the likelihood, employing a surrogate model for the waveform, that is cheaper to evaluate.

There are many possible ways of approaching this problem~\cite{Puerrer2014,Field2014}. Here, we give a brief summary of a procedure for generating a reduced order likelihood and surrogate waveform model. This description follows Field \emph{et al}~\cite{Field2014}, where more details may be found.

As a starting point, we compute $M$ waveform templates $h(t; \boldsymbol{\lambda}_i)$, referred to as the \emph{training set}. The aim is first to find the minimal number $m$ of these waveforms such that all other waveforms in the training set can be well approximated. If the training set is sufficiently dense, the approximation will also be valid for waveforms outside the training set. Once this reduced basis set has been found, the second stage is to identify a set of $m$ discrete times at which it is sufficient to compute the waveform in order to represent it faithfully with the reduced basis. The final stage is to construct a surrogate model that can predict the value of the waveform at those times for arbitrary choices of the model parameters. The algorithm for achieving this is as follows:

\begin{enumerate}
\item Choose the $m$ most differing waveforms and construct an orthonormal basis $\{e_i(t)\}_{i=1}^{i=m}$ from them\footnote{The basis is constructed in a greedy manner. Given an existing basis of size $r<m$, the $(r+1)$th waveform is the member of the training set with the largest residual norm when projected onto the $r$-basis. The first waveform is chosen arbitrarily.}. The corresponding waveform parameters are referred to as \emph{greedy data}. Given the similarity between GWs with different parameters, it is expected that $m \ll M$. Every waveform in the training set, as well as waveforms that are not in the training set, may then be approximated by the expansion
\begin{equation}
h_m(t;\boldsymbol{\lambda}) \approx \sum_{i=1}^{m} c_i(\boldsymbol{\lambda})e_i(t).
\end{equation}
The representation error of the reduced basis (RB) is defined as
\begin{equation}\label{eq:reperror}
\sigma_m \equiv \max_{\boldsymbol{\lambda}}\min_{c_i} \inneri{h(\boldsymbol{\lambda})-h_m(\boldsymbol{\lambda})}{h(\boldsymbol{\lambda})-h_m(\boldsymbol{\lambda})},
\end{equation}
where the minimisation over the coefficients is done by projecting the waveform onto the RB. The value of $m$ is set by a condition on $\sigma_m$; typically we may require that $\sigma_m \lesssim \mathcal{O}(10^{-12})$.

\item Identify $m$ evaluation times $\{T_i\}_{i=1}^{i=m}$ that can be used to construct an empirical interpolant for the RB; these are referred to as \emph{empirical nodes}\footnote{The empirical nodes are selected in a greedy manner. Using $r$ existing nodes and the first $r$ basis functions, an empirical interpolant \eqref{eq:empirical-interpolant} is built for the $(r+1)$th basis element. The $(r+1)$th empirical node is the time at which this interpolant is most different from the actual basis function.}. The goal is to construct an interpolant
\begin{equation}
\label{eq:empirical-interpolant}
\mathcal{I}_m[h](t;\boldsymbol{\lambda}) = \sum_{j=1}^{m}B_j(t)h(T_j;\boldsymbol{\lambda}),
\end{equation}
where
\begin{equation}
B_j(t) \equiv \sum_{i=1}^{m}e_i(t)(V^{-1})_{ij}
\end{equation}
is independent of the system parameters and so may be computed offline. The $V$ matrix is constructed by requiring that $\mathcal{I}_m[h](T_j;\boldsymbol{\lambda}) = h(T_j;\boldsymbol{\lambda})$.

\item At each empirical node, predict the waveform value for arbitrary parameters by fitting $h(T_i;\boldsymbol{\lambda})$ with respect to $\boldsymbol{\lambda}$, using only the greedy data. It is often easier to find fits for the amplitude $A_i$ and phase $\phi_i$ of the waveform independently (as opposed to $h_+$ and $h_x$ individually). The waveform can then be written as
\begin{equation}
h(T_i;\boldsymbol{\lambda}) \approx A_i(\boldsymbol{\lambda})e^{\iu \phi_i(\boldsymbol{\lambda})},
\end{equation}
where the fits are arbitrary functions of the parameters; the specific choice will be determined by the problem at hand.
\end{enumerate}

Once these offline steps have been completed, a surrogate model can be constructed from the empirical interpolant
\begin{equation}
\label{eq:surrogate-model}
h_S(t; \boldsymbol{\lambda}) = \sum_{i=1}^{m}B_i(t)A_i(\boldsymbol{\lambda})e^{\iu \phi_i(\boldsymbol{\lambda})}.
\end{equation}
The model is quick to evaluate at arbitrary $\boldsymbol{\lambda}$ as the $\{B_i\}$ are computed in advance, and the $\{A_i\}$ and $\{\phi_i\}$ are simple analytic expressions, fitted to the numerical data.

\section{\label{sec:likelihood-transform}Likelihood Transform}
A general optimisation problem involves finding the set of parameters that globally maximises some function. Here we will initially calculate the SNR to illustrate the smoothing technique in a detector- and noise-independent way. We then apply the method to the maximisation of a simple noise-dependent likelihood function.

The SNR often has a great deal of structure: there are many local secondary maxima and the true global peak is tall and narrow. This makes the optimisation problem difficult because a high resolution is required and it is possible that any algorithm will stall at a secondary maximum, and thus not explore the full parameter space. It is therefore desirable to be able to sample a smoother distribution that closely mirrors the true SNR surface. To this extent, we follow Wang~\cite{Wang2014} and define a smoothed SNR via a convolution
\begin{equation}
\label{eq:smoothSNR}
\rho_\sigma(\boldsymbol{\lambda}; \boldsymbol{\lambda}_*) \equiv \mathcal{K}_\sigma \star \rho(\boldsymbol{\lambda}; \boldsymbol{\lambda}_*) = \Re[\inneri{h(\boldsymbol{\lambda}_*)}{h_\sigma(\boldsymbol{\lambda})}],
\end{equation}
where $\mathcal{K}_\sigma$ is some smoothing kernel of typical width $\sigma$, we have used the linear property of the SNR and
\begin{equation}
\label{eq:smoothed-h}
h_\sigma(\boldsymbol{\lambda}) \equiv \mathcal{K}_\sigma \star h(\boldsymbol{\lambda}) = \intd{}{}{\mathcal{K}_\sigma(\boldsymbol{\lambda}-\boldsymbol{\lambda}')h(t; \boldsymbol{\lambda}')}{\boldsymbol{\lambda}'}.
\end{equation}

It would be possible to numerically perform this integral, by sampling the waveform space at $k$ points surrounding the desired parameters $\boldsymbol{\lambda}$ and compute the appropriate sum, weighted by $\mathcal{K}_\sigma$. The major drawback to this approach is the required computational time; running time would be increased by $\mathcal{O}(k)$, and in high $N$-dimensional parameter spaces, the required number of points to accurately estimate the integral may be very large, $k \sim \mathcal{O}(2^N)$.

Alternatively, if there exists an analytic expression for the waveform model $h(t; \boldsymbol{\lambda})$, it may be possible to explicitly calculate (or at least approximate) the integral for particular choices of kernel $\mathcal{K}_\sigma$. This would result in an analytic expression for $h_\sigma$, enabling the smoothed SNR to be calculated quickly. This was the approach taken in the previous study of likelihood transform methods~\cite{Wang2014}, where a simple quadratic chirp signal was considered. This approach is highly restrictive since faithful models of likely GW sources are not usually analytic but instead are computed numerically on some parameter grid. The quadratic chirp model used in~\cite{Wang2014} is not a faithful model of any likely GW signal.

We take an approach between these two extremes, showing how the surrogate model \eqref{eq:surrogate-model} can be utilised to simplify the smoothing operation \eqref{eq:smoothed-h} required in likelihood transform techniques\footnote{We later illustrate the technique using an analytic model but, in contrast to Wang~\cite{Wang2014}, our procedure does not rely on this.}. To do this, we must first lose some generality, although the resulting procedure remains general enough to be of wide applicability.

The $N$ as-yet unspecified parameters, $\boldsymbol{\lambda}$, for each waveform are mapped onto the unit cube $\boldsymbol{\theta}$ in parameter space. This allows us to discuss a wide range of waveform models, without being overly specific. We could have equally chosen to map the parameters $\boldsymbol{\lambda}$ onto an infinite or semi-infinite range, giving similar results.

We choose the kernel $\mathcal{K}_\sigma$ to be a multivariate Gaussian with diagonal covariance matrix $\boldsymbol{\Sigma} = \mathrm{diag}(\sigma_1^2,\ldots,\sigma_N^2)$. The smoothed waveform can then be calculated using
\begin{equation}
h_\sigma(\boldsymbol{\theta}) = \intd{}{}{\!}{\boldsymbol{\theta}'} \: h(t; \boldsymbol{\theta}') \prod_{j=1}^{N} \mathcal{N}_j e^{-(\theta_j-\theta'_j)^2/2\sigma_j^2},
\end{equation}
where $\mathcal{N}_j \equiv \mathcal{N}(\theta_j;\sigma_j)$ is a normalisation function that we derive later.

We now make use of the surrogate model \eqref{eq:surrogate-model} to remove the time dependence from the integrals
\begin{align}
\label{eq:gaussian-smoothed-waveform}
h_\sigma(\boldsymbol{\theta}) = \sum_{i=1}^{m}B_i(t)\intd{}{}{\!}{\boldsymbol{\theta}'} &\: A_i(\boldsymbol{\theta}')e^{\iu \phi_i(\boldsymbol{\theta}')} \nonumber\\
&\prod_{j=1}^{N} \mathcal{N}_j e^{-(\theta_j-\theta'_j)^2/2\sigma_j^2}.
\end{align}
The phase of the waveform may be approximated around the evaluation point using a Taylor series
\begin{equation}
\label{eq:phase-approx}
\phi_i(\boldsymbol{\theta}') \approx \phi_i(\boldsymbol{\theta}) + (\boldsymbol{\theta}'-\boldsymbol{\theta}).\nabla\phi_i.
\end{equation}
This is a good approximation when $(\boldsymbol{\theta}'-\boldsymbol{\theta})$ is small, which is true if we choose $\sigma$ to be sufficiently small\footnote{Even for larger values of $\sigma$, the procedure can still be followed. In this case, the smoothed waveform \eqref{eq:phase-shifted-waveform} is not a good approximation to \eqref{eq:gaussian-smoothed-waveform}, but the resulting smoothed likelihood surface may still be sufficiently similar to the unsmoothed surface that MCMC convergence will be accelerated.}. The waveform can then be written as
\begin{align}
\label{eq:phase-shifted-waveform}
h_\sigma(\boldsymbol{\theta}) &= \sum_{i=1}^{m}B_i(t)e^{\iu \phi_i(\boldsymbol{\theta})}\intd{}{}{\!}{\boldsymbol{\theta}'} \: A_i(\boldsymbol{\theta}') \nonumber\\
&\prod_{j=1}^{N} \mathcal{N}_j e^{-(\theta'_j-\theta_j)^2/2\sigma_j^2} \exp\left[\iu (\theta'_j-\theta_j)\partialdiff{\phi_i}{\theta_j}\right].
\end{align}
We are free to choose any set of functions $\{A_i\}$, as long as they accurately fit the numerical waveform data. We first focus on functions that can be decomposed into a short series of separable terms\footnote{If we allow the number of terms to approach infinity, this decomposition can be used to represent any sufficiently smooth function, but the number of required fitting parameters will also approach infinity. By \textit{short} series, we mean that the number of fitting parameters required to represent the function is smaller than the number of elements $m$.}
\begin{equation}
\label{eq:terms-sum}
A_i(\boldsymbol{\theta}) = \sum_{\mathrm{terms}}\prod_{j=1}^{N} f_{j}(\theta_j),
\end{equation}
where $f_{j}$ are arbitrary functions that may in principle be different for each term. We will discuss generic amplitudes later. The smoothed waveform can then be written as
\begin{align}
h_\sigma(\boldsymbol{\theta}) &= \sum_{i=1}^{m}B_i(t)e^{\iu \phi_i(\boldsymbol{\theta})}\sum_{\mathrm{terms}} \intd{}{}{\!}{\boldsymbol{\theta}'} \:\nonumber\\
&\prod_{j=1}^{N} \mathcal{N}_j f_j(\theta'_j)\, e^{-(\theta'_j-\theta_j)^2/2\sigma_j^2} \exp\left[\iu (\theta'_j-\theta_j)\partialdiff{\phi_i}{\theta_j}\right],
\end{align}
which can be simplified to
\begin{equation}
\label{eq:smooth-waveform}
h_\sigma(\boldsymbol{\theta}) = \sum_{i=1}^{m}B_i(t)e^{\iu \phi_i(\boldsymbol{\theta})}\sum_{\mathrm{terms}} \prod_{j=1}^{N} f_j(\theta_j;\sigma_j),
\end{equation}
where we define
\begin{align}
f_j(\theta_j;\sigma_j) &\equiv  \intd{0}{1}{\!}{\theta'_j} \: \mathcal{N}_j \nonumber\\
&f_j(\theta'_j)\, e^{-(\theta'_j-\theta_j)^2/2\sigma_j^2} \exp\left[\iu (\theta'_j-\theta_j)\partialdiff{\phi_i}{\theta_j}\right].
\end{align}
The appeal of such an approach is immediate: the smoothed waveform takes an identical form to the surrogate model, but with the replacement $f_j(\theta_j) \rightarrow f_j(\theta_j; \sigma_j)$.

\subsection{Polynomial amplitudes}
\label{sec:poly}
We now consider the specific case of amplitude functions $A_i(\boldsymbol{\theta})$ that can be well described by polynomials. In this case, all of the $f_j(\theta_j)$ will be powers of $\theta_j$. We are hence interested in integrals of the form
\begin{equation}
f(\theta;\sigma,n) \equiv  \intd{0}{1}{\!}{\theta'} \: \mathcal{N} {\theta'}^{n}\, e^{-(\theta'-\theta)^2/2\sigma^2} \exp\left[\iu (\theta'-\theta)\partialdiff{\phi_i}{\theta}\right],
\end{equation}
where we have dropped the $j$ subscripts, for clarity. Completing the square in the exponential terms leads to
\begin{equation}
\label{eq:polyf}
f(\theta;\sigma,n) =  \mathcal{N} \exp\left[-\frac{\sigma^2}{2}\left(\partialdiff{\phi_i}{\theta}\right)^2\right] C^n(\theta+\iu \sigma^2 \partialdiff{\phi_i}{\theta}; \sigma),
\end{equation}
where we have defined
\begin{equation}
C^n(z;\sigma) \equiv  \intd{0}{1}{\!}{z'} \: {z'}^n\, e^{-(z'-z)^2/2\sigma^2}.
\end{equation}
Integrating this by parts gives the recurrence relation
\begin{widetext}
\begin{align}
C^n(z;\sigma) &=  \frac1{(n+1)}e^{-(1-z)^2/2\sigma^2} + \frac1{\sigma^2(n+1)}\intd{0}{1}{\!}{z'} \: \left\{{z'}^{n+2} - z{z'}^{n+1}\right\}\, e^{-(z'-z)^2/2\sigma^2},\\
\label{eq:Cn-recurrence2}
&= \frac1{(n+1)}e^{-(1-z)^2/2\sigma^2} + \frac1{\sigma^2(n+1)}\left\{C^{n+2}(z;\sigma) - zC^{n+1}(z;\sigma)\right\},\\
&= z C^{n-1}(z;\sigma) + \sigma^2\left((n-1)C^{n-2}(z;\sigma) - e^{-(1-z)^2/2\sigma^2}\right),
\end{align}
\end{widetext}
where the last line follows from relabelling $n$ by $n-2$ in \eqref{eq:Cn-recurrence2}. We can perform the integral explicitly for the first two terms:
\begin{equation}
\label{eq:C0}
C^0(z;\sigma) = \sqrt{\frac{\pi}{2}}\sigma\left(\erf\left(\frac{1-z}{\sqrt{2}\sigma}\right)+\erf\left(\frac{z}{\sqrt{2}\sigma}\right)\right),
\end{equation}
\begin{equation}
C^1(z;\sigma) = z C^0(z;\sigma) + \sigma^2\left(e^{-z^2/2\sigma^2}-e^{-(1-z)^2/2\sigma^2}\right).
\end{equation}
To correctly normalise the Gaussian kernel, we must set $\mathcal{N}_j = 1/C^0(\theta_j;\sigma_j)$. To construct a smoothed waveform, we write out the surrogate model, including the polynomial fit for the amplitude functions, and then make the replacements\footnote{In particular, any term that does not depend on $\theta_j$ should be multiplied by $f(\theta_j;\sigma_j,0)$.}
\begin{equation}
\label{eq:poly-smooth-sub}
\theta_j^n \rightarrow f(\theta_j;\sigma_j,n).
\end{equation}

We note the expected property that $f(\theta;\sigma=0,n) = \theta^n$, thus recovering the original polynomial in the case of zero smoothing.

\subsection{Fourier amplitudes}
Rather than polynomials, it may be desirable to decompose the amplitude functions into Fourier components
\begin{equation}
\label{eq:fourier-series}
A(\boldsymbol{\theta}) = \sum_{k_1,k_2,\ldots k_N}A_{\boldsymbol{k}} e^{2\pi \iu \boldsymbol{k}.\boldsymbol{\theta}},
\end{equation}
where we have dropped the $i$ subscript for clarity, $\boldsymbol{k} = \{k_1,k_2,\ldots k_N\}$, the sum runs over positive and negative integers, and
\begin{equation}
A_{\boldsymbol{k}} = \intd{0}{1}{A(\boldsymbol{\theta})e^{-2\pi \iu \boldsymbol{k}.\boldsymbol{\theta}}}{\boldsymbol{\theta}}.
\end{equation}
We note that \eqref{eq:fourier-series} is of the form of \eqref{eq:terms-sum} and so we are interested in computing integrals of the form
\begin{align}
f(\theta;\sigma,n,k) &\equiv  \intd{0}{1}{\!}{\theta'} \: \mathcal{N}\nonumber\\
&{\theta'}^{n}\, e^{\iu k\theta'}\, e^{-(\theta'-\theta)^2/2\sigma^2} \exp\left[\iu (\theta'-\theta)\partialdiff{\phi_i}{\theta}\right],
\end{align}
where we have also included a polynomial factor, for generality. Following the same procedure as before leads to
\begin{align}
f(\theta;\sigma,n,k) =  \mathcal{N} \exp&\left[-\frac{\sigma^2}{2}\left(\partialdiff{\phi_i}{\theta} + k\right)^2 + \iu k\theta\right]\nonumber\\
&C^n\left(\theta+\iu \sigma^2 \left(\partialdiff{\phi_i}{\theta} + k\right); \sigma\right),
\end{align}
where the calculation of $C^n$ is discussed in the previous section. These can then be used in \eqref{eq:fourier-series}, replacing each exponential factor according to
\begin{equation}
e^{2\pi \iu k\theta} \rightarrow f(\theta;\sigma,0,2\pi k).
\end{equation}

\subsection{\label{sec:generic-amplitudes}Generic amplitudes}
In some cases, it may not be possible to decompose the amplitude functions into a series of appropriate separable functions. In this case, we may use a Taylor series to approximate
\begin{equation}
A_i(\boldsymbol{\theta}') \approx A_i(\boldsymbol{\theta}) + (\boldsymbol{\theta}'-\boldsymbol{\theta}).\nabla A_i.
\end{equation}
As with the Taylor series in the phase \eqref{eq:phase-approx}, this is a good approximation for sufficiently small smoothing widths $\sigma$. Substituting it into \eqref{eq:phase-shifted-waveform} results in
\begin{widetext}
\begin{align}
h_\sigma(\boldsymbol{\theta}) = \sum_{i=1}^{m}B_i(t)e^{\iu \phi_i(\boldsymbol{\theta})} \bigg\{& A_i(\boldsymbol{\theta}) \prod_{j=1}^{N} f(\theta_j;\sigma_j,0)\nonumber\\
 &+ \sum_{k=1}^{N} \intd{0}{1}{\!}{\theta'_k} \mathcal{N}_k \left(\theta'_k-\theta_k\right) \partialdiff{A_i}{\theta_k} e^{-(\theta'_k-\theta_k)^2/2\sigma_k^2} \exp\left[\iu (\theta'_k-\theta_k)\partialdiff{\phi_i}{\theta_k}\right]\prod_{j\neq k} f(\theta_j;\sigma_j,0)\bigg\},
\end{align}
which can be written as
\begin{equation}
\label{eq:generic-smoothed-waveform}
h_\sigma(\boldsymbol{\theta})=\sum_{i=1}^{m}B_i(t)e^{\iu \phi_i(\boldsymbol{\theta})} \bigg\{A_i(\boldsymbol{\theta}) + \sum_{k=1}^{N} \partialdiff{A_i}{\theta_k}\frac{f(\theta_k;\sigma_k,1)-\theta_k f(\theta_k;\sigma_k,0)}{f(\theta_k;\sigma_k,0)}\bigg\}\prod_{j=1}^{N} f(\theta_j;\sigma_j,0),
\end{equation}
\end{widetext}
where $f(\theta;\sigma,n)$ is defined in \eqref{eq:polyf}.

With these results, it should be possible to produce a smoothed SNR or likelihood for any given surrogate model waveform. We note that the smoothed waveforms do not necessarily have to be accurate (neglected terms of higher order in $A_i$ and $\phi_i$ may be large) as long as the resulting smoothed surface displays the desired properties that the global maximum is broadened and the number of secondary maxima has been reduced.

Using \eqref{eq:generic-smoothed-waveform}, the expected additional computational cost of evaluating a smoothed waveform with $N$ parameters is $\mathcal{O}(N)$, in comparison to $\mathcal{O}(2^N)$ for naive likelihood smoothing.

\section{\label{sec:examples}Chirp Waveforms}
As an illustration of our method, we consider the gravitational waveforms expected from a circular compact binary. Such waveforms are well-approximated by a high-order post-Newtonian (PN) expansion~\cite{Blanchet2014} in the dimensionless variable $x = (G M \Omega/c^3)^{2/3}$, where $M$ is the total mass of the system and $\Omega = \dot{\Phi}$ is the angular frequency of the binary, computed as a time derivative of the binary phase $\Phi$. The expansion takes the form
\begin{equation}
h_{+,\times} = \frac{2G\mu x}{c^2 R}\sum_{p=0}^{\infty}x^{p/2}\underset{p/2}{H} \!_{+,\times}(\psi,\iota;\ln x) + \mathcal{O}(R^{-2}),
\end{equation}
where $\mu$ is the reduced mass, $R$ is the distance to the source, $\iota$ is the inclination of the binary and $\psi$ is the binary phase distorted by tails
\begin{equation}
\psi = \Phi - \frac{2G M_{\mathrm{ADM}}}{c^3}\Omega\ln\left(\frac{\Omega}{\Omega_0}\right).
\end{equation}
The tail integrals are a result of the nonlinear interaction between the source and the emitted GWs~\cite{Blanchet1993}. Here one must use the binary's mass monopole $M_{\mathrm{ADM}}$, which includes all contributions to the mass-energy of the binary. At 1PN order, it may be computed as~\cite{Blanchet2008}
\begin{equation}
M_\mathrm{ADM} = M\left(1-\frac{\nu}{2}x\right),
\end{equation}
where $\nu \equiv \mu/M$ is the symmetric mass ratio, which takes values between $0$ (test particle limit) and $1/4$ (equal mass). $\Omega_0$ is often chosen to be the lower cutoff of the detector band; we make the choice $\Omega_0 =\nobreak 10\pi \, \mathrm{rad \, s^{-1}}$.

To demonstrate the approach, we now make some simplifying assumptions. First, the expansion will be truncated at finite PN order: we specify the phase $\Phi$ up to 3.5PN (this is given explicitly in appendix \ref{app:PNphase}); the amplitude to 2PN; and $M_\mathrm{ADM}$ to 1PN\footnote{It is not a requirement to have consistency in PN orders between the amplitude and phase.}. The logarithmic term in $\psi$ is then at 4PN order relative to the dominant phase contribution, and is included for completeness.  Secondly, we will consider aligned binaries that have $\iota = 0$. In this case, the relevant expansion functions for the plus polarisation are all proportional to $\cos(2\psi)$ and the cross terms are proportional to $\sin(2\psi)$. We can therefore write a two-parameter family of complex PN waveforms as
\begin{equation}
h(\mathcal{M},\nu;t) = \frac{2G\mathcal{M}\nu^{2/5}}{c^2 R}x(t) H(x(t),\nu) e^{\iu(2 \psi(t)+\pi)},
\end{equation}
where $\mathcal{M} = \mu^{3/5}M^{2/5}$ is the chirp mass and the amplitude function $H$ takes the form~\cite{Blanchet1996}
\begin{align}
H(x,\nu) = 2 + &\frac1{3}\left(\nu-13\right)x + 4\pi x^{3/2}\nonumber\\
  &+ \frac1{180}\left(15\nu^2 - 635\nu - 837\right)x^2.
\end{align}
The clean separation into an amplitude and phase is a consequence of our simplifying assumptions, but this is not a necessary requirement for the smoothing procedure.

\subsection{Newtonian chirp}\label{sec:0pn-chirp}
We first simplify the waveform model even more, truncating at 0PN order so that the waveform has only one free intrinsic parameter, the chirp mass $\mathcal{M}$. We choose $t_c = 0.4\mathrm{s}$, and set a fiducial distance of $R=1\mathrm{Mpc}$, although this is unimportant as we normalise the waveforms to unity. We shall consider chirp masses in the range $1 \leq \mathcal{M}/M_\odot \leq 20$, corresponding roughly to the range considered in recent LIGO searches~\cite{Abbott2009a}, and inject a signal with $\mathcal{M}_* = 2.2 M_\odot$. 

We can calculate the FIM for this waveform model at our injection parameters. Using the unit parameter $\theta = (\mathcal{M}/M_\odot-1)/19$, we find that the global maximum of the likelihood has a characteristic scale of $7\times 10^{-4}$. 

\subsubsection{Idealised data}
\begin{figure}[htbp]
\centering
\includegraphics[width=0.45\textwidth]{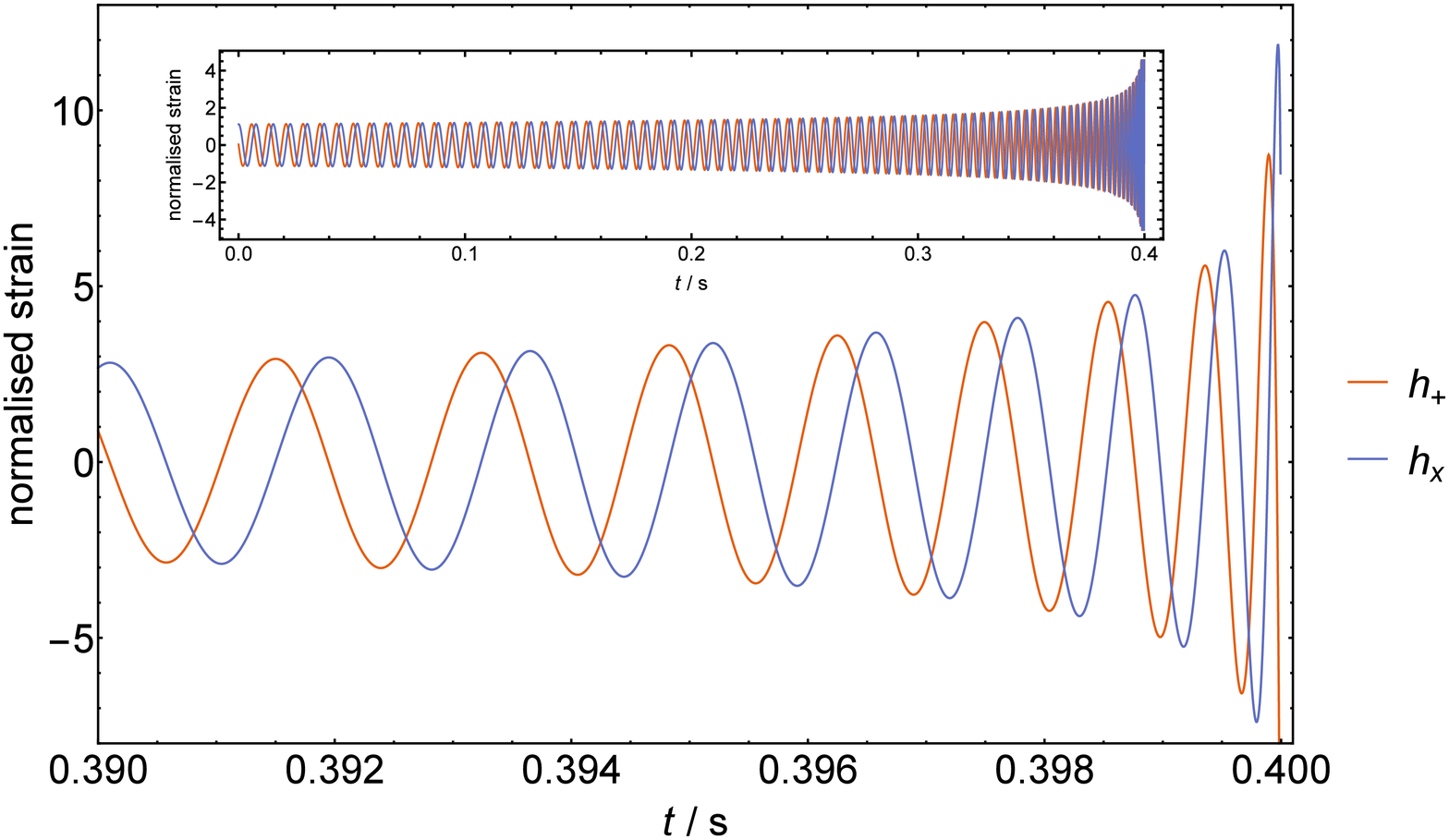}
\caption{\label{fig:injected-waveform}Gravitational waveform from a compact binary with chirp mass $\mathcal{M}=2.2M_\odot$ in the final $0.01\mathrm{s}$ before coalescence. Inset is the same waveform for a period of $0.4\mathrm{s}$ before coalescence. The amplitude is set by the requirement that $\inneri{h}{h}=1$.}
\end{figure}

We initially perform calculations without noise, to demonstrate the underlying properties of the smoothing technique. Our injected waveform is plotted in figure \ref{fig:injected-waveform}.

\begin{figure}[htbp]
\centering
\includegraphics[width=0.45\textwidth]{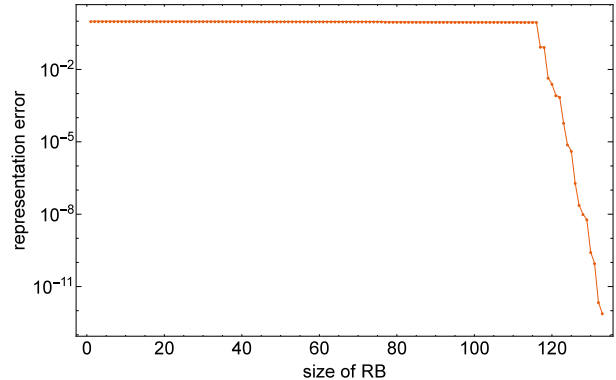}
\caption{\label{fig:RB-error}The representation error of our RB for the 0PN waveform family as a function of the number of elements in the basis.}
\end{figure}

We construct a surrogate model for our waveform family, as discussed in section \ref{sec:surrogate}, using a training set of 501 waveforms, sampled at a frequency $f_\mathrm{sample} = 20 \,\mathrm{kHz}$ and with chirp masses selected such that they sample the frequency $\Omega$ uniformly\footnote{Selecting a training set with uniform values of $\mathcal{M}$ resulted in a poor surrogate model for low values of $\mathcal{M}$ (high frequencies).}. We target a representation error of $10^{-12}$ and the resulting RB contains 133 elements. The error as a function of the size of the RB is plotted in figure \ref{fig:RB-error}.

To check the faithfulness of the RB, we compute the representation error \eqref{eq:reperror} for 1000 waveforms \emph{not} in the training set, generated with random values of $\mathcal{M}$ sampled uniformly from the allowed range; all of the waveforms had $\sigma < 10^{-12}$, with typical values of $\sigma \sim 10^{-15}$.

\begin{figure}[htbp]
\centering
\includegraphics[width=0.45\textwidth]{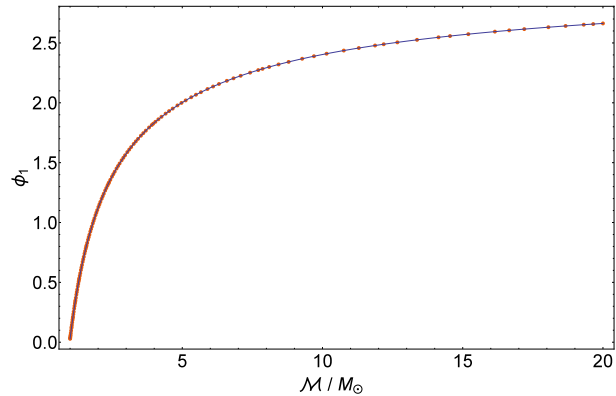}
\caption{\label{fig:phi1-fit}The phase of the gravitational waveform, evaluated at the first empirical node, as a function of chirp mass. The dots show the values of the phase evaluated at the greedy points selected while constructing the RB. The line is a fit to the data, which in this case is exact.}
\end{figure}

Using the training set waveforms identified in generating the RB, we perform a fit to the waveform phase at each empirical node, using the functional form
\begin{equation}
\phi_i(\mathcal{M}) = a_1 + a_2\mathcal{M}^{a_3},
\end{equation}
where $\{a_k\}_{k=1}^{3}$ are the fitting parameters. Figure \ref{fig:phi1-fit} shows the phase as a function of chirp mass, evaluated at the first empirical node, along with the fit; the good agreement is not surprising as the fitting function was motivated by our knowledge of the exact waveform model. Similar results are found for the phase at the other empirical nodes.

\begin{figure}[htbp]
\centering
\includegraphics[width=0.45\textwidth]{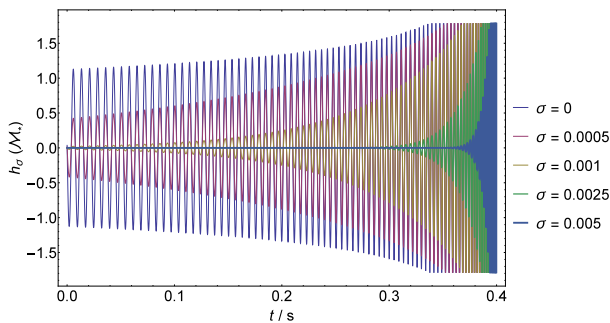}
\caption{\label{fig:smoothed-waveforms}The smoothed waveform \eqref{eq:smooth-waveform}, computed for our injection parameters with a chirp mass $\mathcal{M} = 2.2 M_\odot$. Thicker lines correspond to larger values of the smoothing width $\sigma$. The $\sigma = 0$ waveform is identical to that in figure \ref{fig:injected-waveform}. }
\end{figure}

For the 0PN waveforms, the amplitude at each empirical node is simply a constant
\begin{equation}
A_i(\mathcal{M}) = b_i.
\end{equation}
Following the procedure in section \ref{sec:poly}, we then make the replacement at each empirical node
\begin{equation}
b_i \rightarrow b_i\, \mathcal{N} \exp\left[-\frac{\sigma^2}{2}\left(\partialdiff{\phi_i}{\theta}\right)^2\right] C^0(\theta+\iu \sigma^2 \partialdiff{\phi_i}{\theta}; \sigma),
\end{equation}
where $\theta = (\mathcal{M}/M_\odot-1)/19$ is the chirp mass mapped onto the unit cube. The resulting smoothed waveforms for our injection signal for different values of $\sigma$ are shown in figure \ref{fig:smoothed-waveforms}.

\begin{figure}[htbp]
\centering
\includegraphics[width=0.45\textwidth]{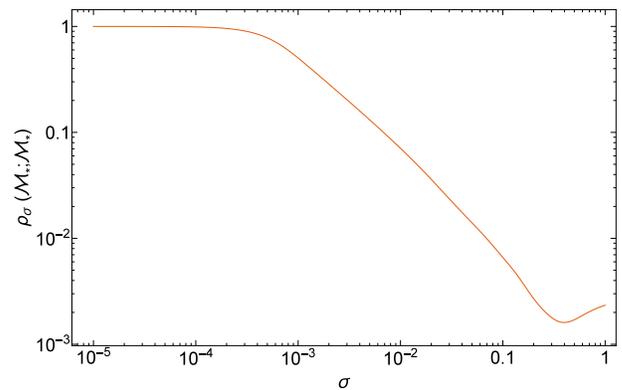}
\caption{\label{fig:smoothed-overlaps}The smoothed SNR \eqref{eq:smoothSNR}, computed for our injection parameters with a chirp mass $\mathcal{M}_* = 2.2 M_\odot$, as a function of smoothing width.}
\end{figure}

The choice of $\sigma$ is arbitrary, but clearly has a large impact on the resulting performance of the algorithm: choose $\sigma$ too small and the smoothed waveforms will be indistinguishable from the unsmoothed waveforms; choose $\sigma$ too large and the smoothed waveforms will be sufficiently dissimilar that the transform technique will not aid in locating the global maximum. To get some idea of the required scale, we compute the smoothed SNR $\rho_\sigma(\mathcal{M}_*;\mathcal{M}_*)$ of our injection waveform, as a function of smoothing width, shown in figure \ref{fig:smoothed-overlaps}. We see that choosing values $\sigma \sim \mathcal{O}(10^{-3})$ results in smoothed waveforms that are different to the unsmoothed waveform, but which still give a high SNR $\rho_\sigma \gtrsim 0.1$. This is consistent with the typical scale of the peak of the likelihood distribution obtained from the FIM.

\begin{figure}[htbp]
\centering
\includegraphics[width=0.45\textwidth]{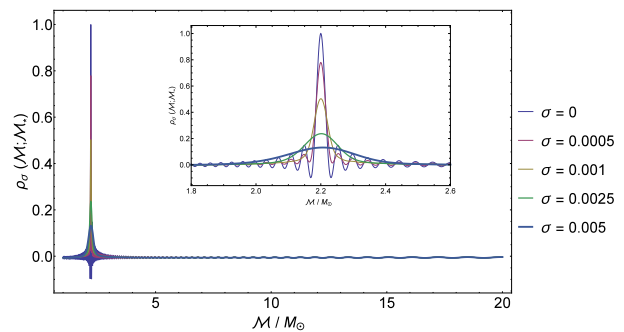}
\caption{\label{fig:smoothed-overlap-curves}The smoothed SNRs \eqref{eq:smoothSNR}, computed across the allowed range of chirp masses, for a selection of smoothing widths. The inset plot shows the behaviour around the injection value $\mathcal{M}_* = 2.2 M_\odot$. Thicker lines correspond to larger values of $\sigma$.}
\end{figure}

To simulate searching for the global maximum, we compute the smoothed SNR $\rho_\sigma(\mathcal{M}; \mathcal{M}_*)$ from \eqref{eq:smoothSNR} on a grid of $\{\sigma, \mathcal{M}\}$ values; the resulting curves for different values of $\sigma$ are shown in figure \ref{fig:smoothed-overlap-curves}. It can be seen that the desired smoothing properties have been achieved: the global peak has been broadened and the number of secondary maxima has been reduced.

If the likelihood transform is to be useful, the time taken to evaluate $\rho_\sigma$ must not greatly exceed that taken to evaluate $\rho_0$\footnote{We compare to $\rho_0$ rather than $\rho$ to make use of the pre-calculated surrogate model.}. In this example, we found that to compute the smoothed SNR took roughly a factor of 4 longer than $\rho_0$. The time taken to calculate the training set data and to produce the surrogate model is relatively large, but this can be done offline.

\subsubsection{Noisy data}
We now consider analysing a data set containing an injected signal and Gaussian white noise. We use the plus polarisation of the waveform only, which is equivalent to having an optimally oriented source observed with a right-angle detector aligned with the principal polarisation axes of the system, as discussed in Section~\ref{sec:GWs}. The SNR of the injected signal is approximately $400$. This is a particularly large SNR, albeit not unusual for, say, supermassive black hole mergers observed with space-based detectors. We deliberately chose this value to make the problem of secondary maxima more pronounced in our example. If the likelihood transform approach can accelerate convergence in this kind of problem it will readily solve the same problems when the SNR is lower.

\begin{figure}[htbp]
\centering
\includegraphics[width=0.45\textwidth]{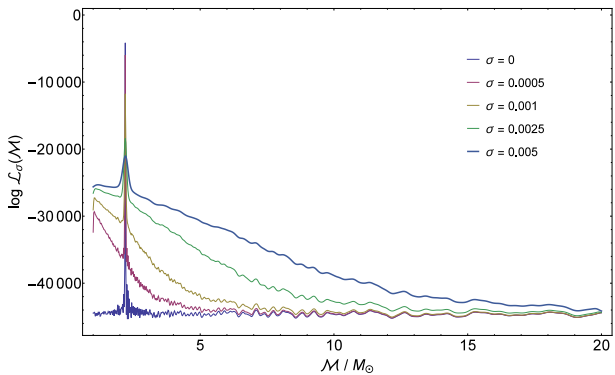}
\caption{\label{fig:smoothed-likelihood-curves}The likelihood function computed using the smoothed waveform model for different values of the smoothing width. Thicker lines correspond to larger values of $\sigma$.}
\end{figure}

For this one-dimensional example, it is possible to calculate the likelihood on a grid of parameter values, using both the unsmoothed and smoothed waveform models. Figure \ref{fig:smoothed-likelihood-curves} shows the computed likelihood for different smoothing widths. The desired smoothing properties are apparent: the global peak has broadened but remains close to the true value, and the number of secondary maxima has been reduced.

We now illustrate the more realistic situation of performing an MCMC search on the posterior distribution. We use a uniform prior on the chirp mass, and a Gaussian proposal distribution of width $10^{-3}$. We choose $500$ seeds $\mathcal{M}_\mathrm{seed}$ drawn from a uniform distribution across the allowed range of chirp masses, and start different MCMCs from each value: one using the unsmoothed surrogate waveform models; and four others using the smoothed model with different values of $\sigma$.

We run the chains for a small number of Metropolis-Hastings steps: 2500 for the unsmoothed chain and 1000 for the smoothed chains. We do not expect the chains to have converged on a stationary distribution by this point; it instead gives an indication of how quickly convergence may occur. The chain lengths are chosen such that the computational time is roughly equal for each type of chain.

\begin{figure}[htbp]
\centering
\includegraphics[width=0.45\textwidth]{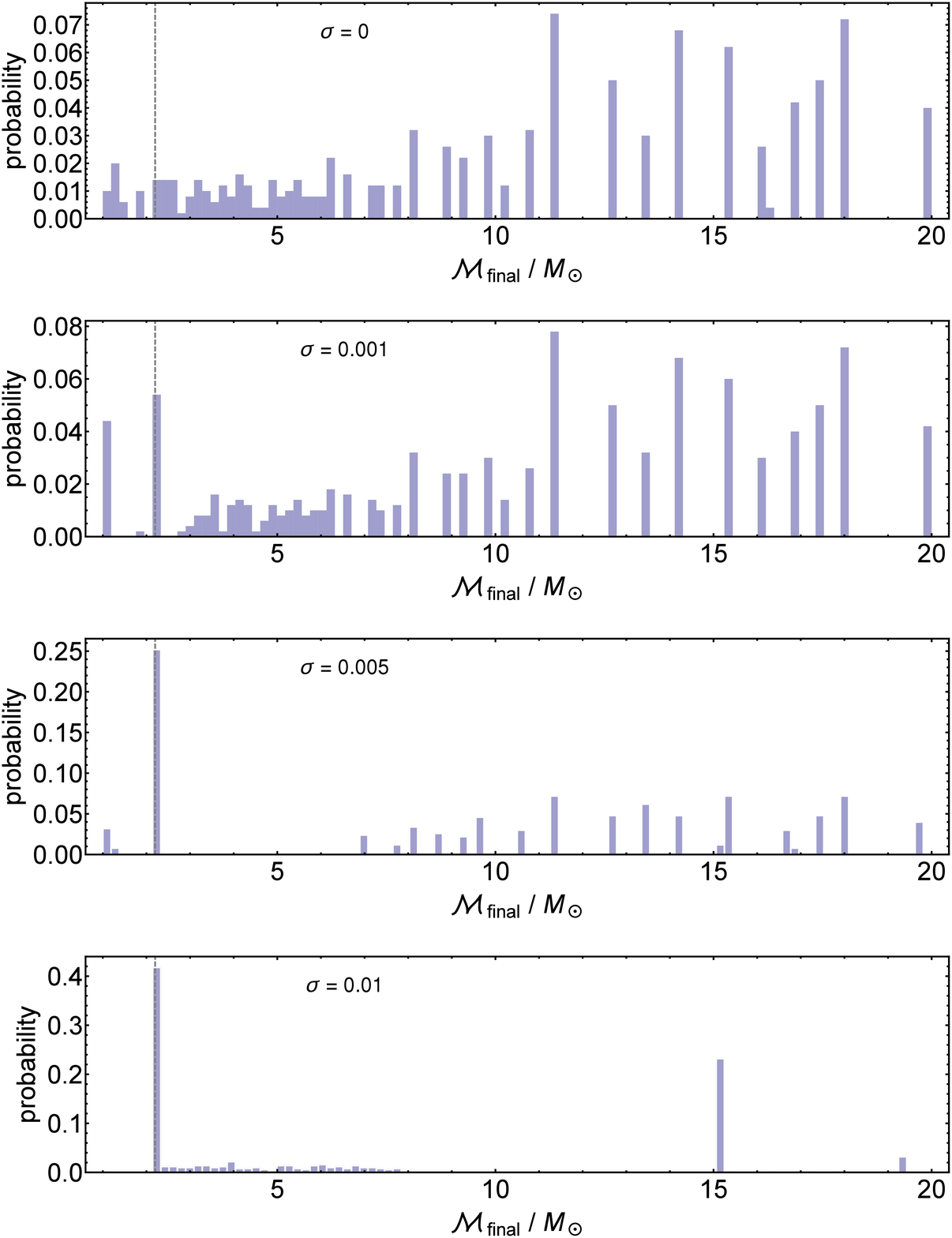}
\caption{\label{fig:short-MCMC-finalvals}Distribution of final chirp masses after a short MCMC, using likelihood functions with different amounts of smoothing. The narrowness of the peaks is indicative that the chains have located local maxima. The vertical grey dashed line is positioned at the true value $\mathcal{M}_* = 2.2 M_\odot$.}
\end{figure}

\begin{figure}[htbp]
\centering
\includegraphics[width=0.45\textwidth]{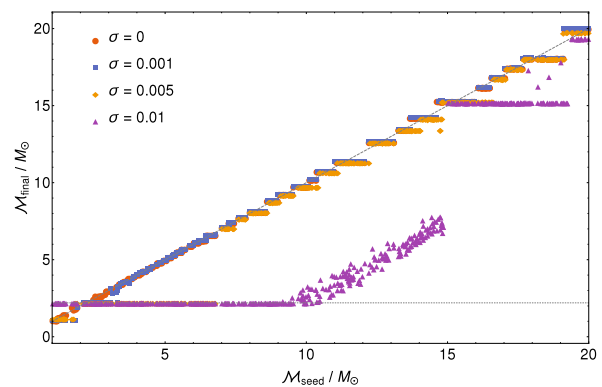}
\caption{\label{fig:short-MCMC-final-vs-seed}The final chirp mass, as a function of the seed mass, for MCMCs using different smoothed likelihood functions. The dashed sloped line is $\mathcal{M}_\mathrm{final} = \mathcal{M}_\mathrm{seed}$, indicating chains that did not move far from their starting point. The dotted line $\mathcal{M}_\mathrm{final} = \mathcal{M}_*$ denotes the true value.}
\end{figure}

The distributions of final chirp masses for both the unsmoothed and smoothed cases are shown in figure \figref{short-MCMC-finalvals}; the unsmoothed chains locate local maxima close to their seed value and so the final distribution is roughly uniform. On the other hand, the smoothed cases show that a significant number of chains have ended up near the global maximum at $\mathcal{M} = 2.2 M_\odot$. The large peak visible at $\mathcal{M}_\mathrm{final} \approx 15 M_\odot$ is a local maximum that has accrued many chains within the small number of steps that we have run; as the number of steps is increased we expect these chains to move towards the global peak, as illustrated by the chains in the vicinity of $5 M_\odot$. Figure \figref{short-MCMC-final-vs-seed} shows the final chain values as a function of the seed chirp mass. Systems that start close to the global maximum locate it quickly using the smoothed waveform model.

\begin{figure}[htbp]
\centering
\includegraphics[width=0.45\textwidth]{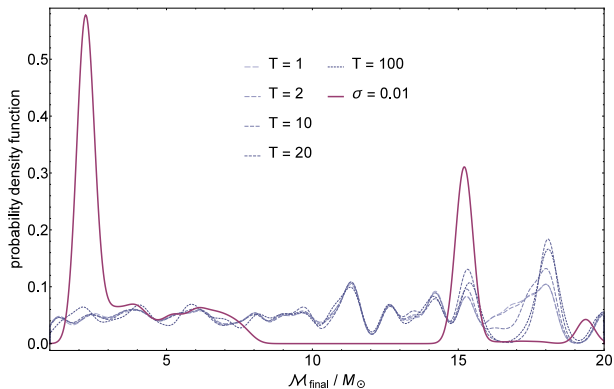}
\caption{\label{fig:short-MCMC-finalvals-withT}Distribution of final chirp masses for chains at different temperatures (dashed lines), compared to the smoothed chain (solid line).}
\end{figure}

It is also possible to perform a comparison with simulated annealing and parallel tempering techniques. We run MCMCs for 2500 steps on the unsmoothed likelihood but at different temperatures, starting at the same chirp mass seed values as above. The final distributions of chirp masses are shown in figure \figref{short-MCMC-finalvals-withT}. Changing the temperature of the chain does help to explore the parameter space, but is not as effective as likelihood smoothing at locating the largest peaks.

\subsection{Higher order post-Newtonian chirp}
We now consider our full PN waveform (up to 3.5PN in the phase and 2PN in the amplitude), with the two mass parameters $\mathcal{M}$ and $\nu$ allowed to vary. We look at chirp masses in the range $1 \leq \mathcal{M}/M_\odot \leq 20$ and mass ratios in the range\footnote{If the component masses are $10 M_\odot$ and $1.4 M_\odot$, corresponding to a fiducial black hole-neutron star binary, the symmetric mass ratio is $\nu = 0.107725$.} $0.1 \leq \nu \leq 0.25$.

\subsubsection{Idealised data}
\begin{figure}[htbp]
\centering
\includegraphics[width=0.45\textwidth]{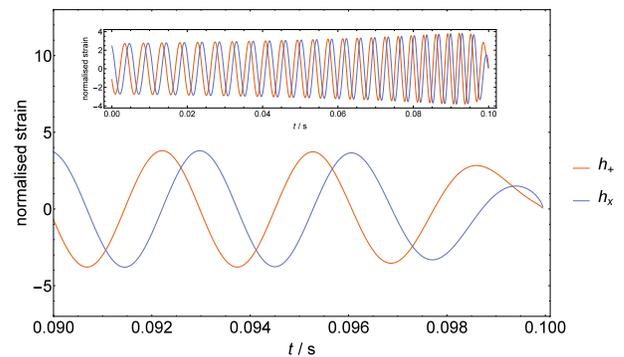}
\caption{\label{fig:HighPN-injected-waveform}PN gravitational waveform from a compact binary with chirp mass $\mathcal{M}=2.2M_\odot$ and symmetric mass ratio $\nu = 0.18$ in the final $0.01\mathrm{s}$ before coalescence. Inset is the same waveform for a period of $0.1\mathrm{s}$ before coalescence. The amplitude is set by the requirement that $\inneri{h}{h}=1$.}
\end{figure}

For an initial study, we look at waveforms in the absence of noise. Our injected waveform has parameters $\mathcal{M}_* = 2.2 M_\odot$, $\nu_*= 0.18$, $t_c = 0.1\mathrm{s}$ and $R=1\mathrm{Mpc}$, and is plotted in figure \figref{HighPN-injected-waveform}. The typical scales of variation about the true values obtained from the FIM are $2\times10^{-3}$ for $\theta_1$ (unit chirp mass) and $0.12$ for $\theta_2$ (unit symmetric mass ratio).

\begin{figure}[htbp]
\centering
\includegraphics[width=0.45\textwidth]{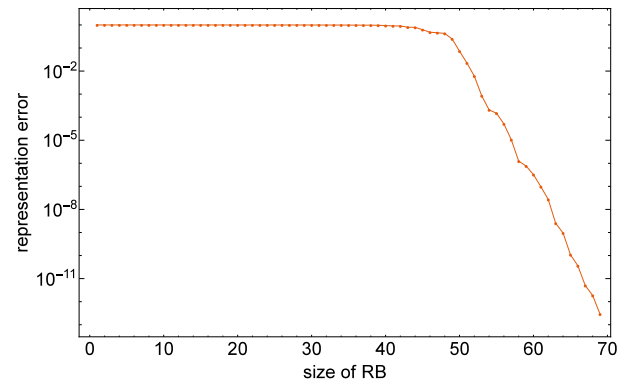}
\caption{\label{fig:HighPN-RB-error}The representation error of the RB for the PN waveform family as a function of the number of elements in the basis.}
\end{figure}

We construct a surrogate model for the two-dimensional PN waveform family using a training set of $190\times 190$ waveforms, sampled at a frequency $f_\mathrm{sample} = 20 \,\mathrm{kHz}$. The chirp masses for the training set waveforms are the same as those in section \ref{sec:0pn-chirp}, while $\nu$ is selected uniformly from the allowed range for each value of $\mathcal{M}$. We target a representation error of $10^{-12}$ and the resulting RB contains 69 elements\footnote{This number should not be directly compared to the 133 RB elements in section \ref{sec:0pn-chirp} since the waveform models are different.}. The error as a function of the size of the RB is plotted in figure \figref{HighPN-RB-error}. We confirm the faithfulness of the RB by computing the representation error for $2500$ waveforms drawn from a uniform distribution on the parameter space; no error exceeded $10^{-12}$ and typical errors were $\mathcal{O}(10^{-14})$.

Following the surrogate model procedure, we use the training set waveforms identified in generating the RB to perform a fit to the waveform phase at each empirical node. In this work, we are not interested in the efficacy of surrogate models, but instead on their application to likelihood transform methods. To obtain an accurate surrogate model, we therefore use a fitting function based on the exact binary phase \eqref{eq:PNphase}. As a result, the fits are exact to within numerical precision.

The amplitude at each empirical node is now some complicated function of $\mathcal{M}$ and $\nu$. To approximate this at arbitrary values, we perform plane interpolation between the greedy data points, which form an unstructured grid on parameter space. This seems somewhat less than ideal, however, the resulting surrogate model is sufficiently good due to our accurate phase model as well as the fact that the greedy points chosen by the RB algorithm are in important regions of parameter space.

\begin{figure}[htbp]
\centering
\includegraphics[width=0.45\textwidth]{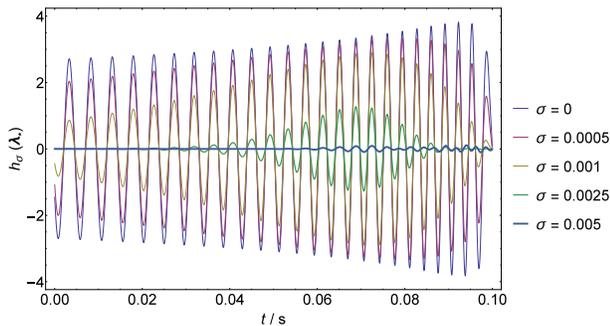}
\caption{\label{fig:HighPN-smoothed-waveforms}The smoothed PN waveform, computed for our injection parameters with a chirp mass $\mathcal{M} = 2.2 M_\odot$ and symmetric mass ratio $\nu = 0.18$. Thicker lines correspond to larger values of the smoothing width $\sigma$. The $\sigma = 0$ waveform is identical to that in figure \ref{fig:HighPN-injected-waveform}. }
\end{figure}

We follow the approach of section \ref{sec:generic-amplitudes} and define a smoothed waveform model according to \eqref{eq:generic-smoothed-waveform}. We choose smoothing widths such that $\sigma_2 = 57 \sigma_1$, consistent with the FIM estimates. The resulting smoothed waveforms for our injection signal for different values of $\sigma \equiv \sigma_1$ are shown in figure \ref{fig:HighPN-smoothed-waveforms}.

\begin{figure}[htbp]
\centering
\includegraphics[width=0.45\textwidth]{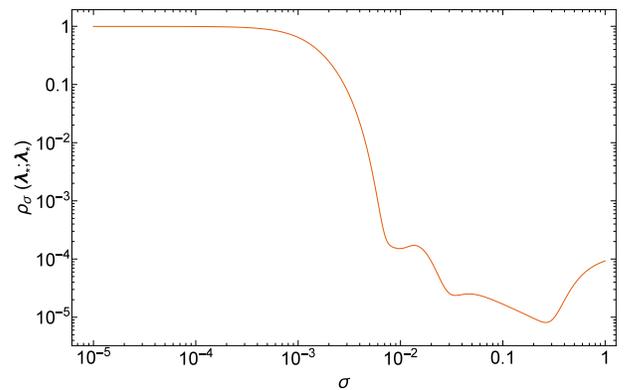}
\caption{\label{fig:HighPN-smoothed-overlaps}The smoothed SNR for the PN waveform family, computed for our injection parameters with a chirp mass $\mathcal{M}_* = 2.2 M_\odot$ and symmetric mass ratio $\nu_* = 0.18$, as a function of smoothing width.}
\end{figure}

The value of $\rho_\sigma(\boldsymbol{\lambda}_*; \boldsymbol{\lambda}_*)$ gives an indication of typical scales on which $\sigma$ is likely to have the desired effect. Figure \ref{fig:HighPN-smoothed-overlaps} shows the smoothed SNR as a function of $\sigma$; widths around $10^{-3}$ give a sufficient level of smoothing, consistent with the FIM. The strange behaviour at larger values of $\sigma$ is due to the peak of the SNR distribution moving away from the true value $\boldsymbol{\lambda}_*$.

\subsubsection{Noisy data}

\begin{figure*}[htbp]
\centering
\includegraphics[width=0.9\textwidth]{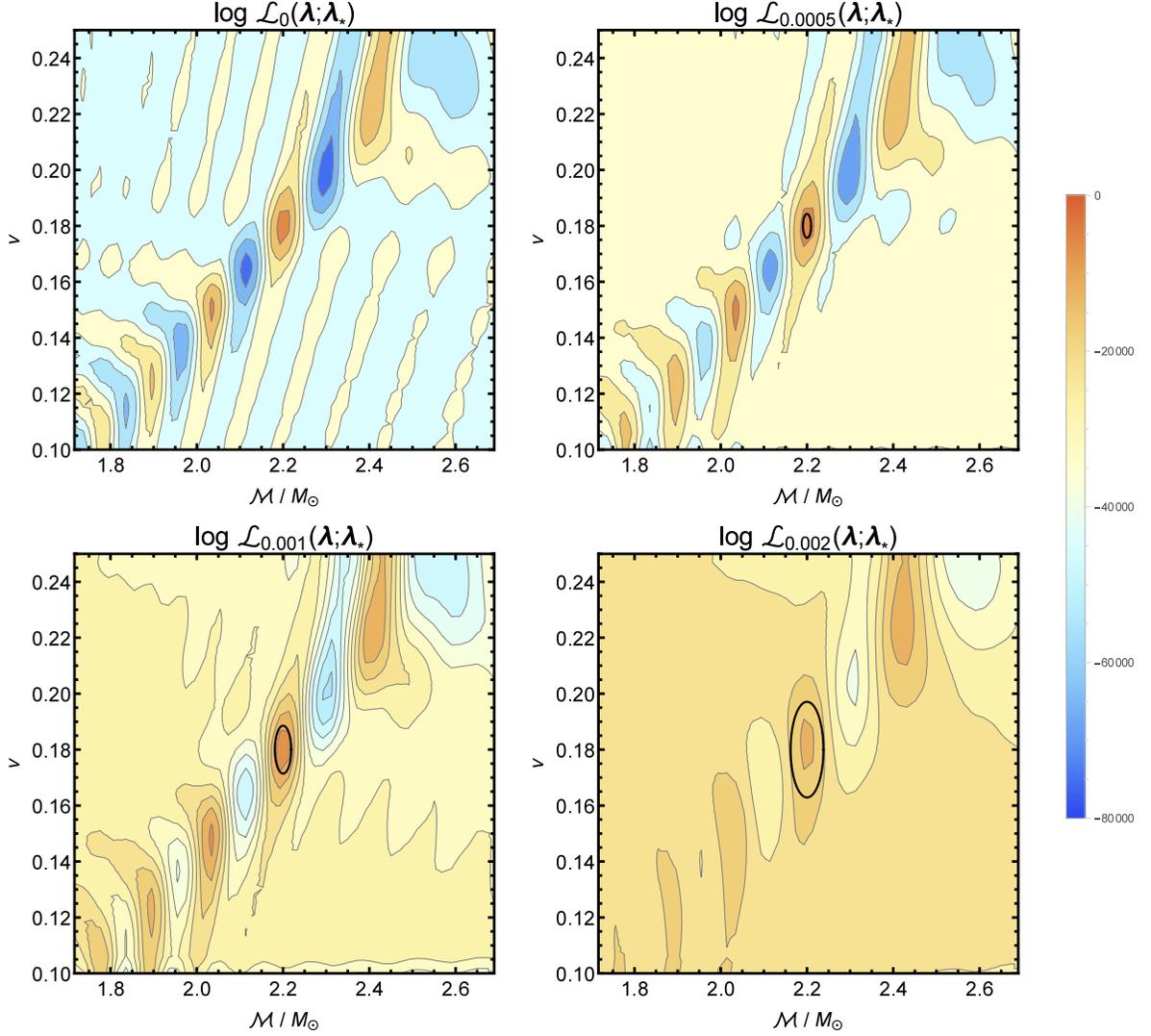}
\caption{\label{fig:HighPN-smoothed-likelihood-curves}The smoothed log-likelihoods for our PN waveform family using a selection of smoothing widths. The black ellipses are centred on the injection values and are sized according to the smoothing width.}
\end{figure*}

As with the 0PN waveform, we now consider the plus polarisation of our injection signal in the presence of white noise. With just two parameters, it is still feasible to map out the likelihood surface for different amounts of smoothing, shown in figure \figref{HighPN-smoothed-likelihood-curves} along with an indication of the size of the smoothing region. As the value of $\sigma$ is increased, we see a decrease in the amount of structure on the likelihood surface.

\section{\label{sec:conclusions}Conclusions}
We have devised a practical scheme for using gravitational wave surrogate models to perform likelihood transform techniques, and have demonstrated that this can be used to accelerate convergence of MCMC methods. The advantage of smoothing the likelihood surface, rather than simply rescaling it (as is done in simulated annealing), is that the number of secondary maxima is reduced. The convolution required to perform this smoothing is expensive to evaluate numerically as it would require the generation of many waveform models. We make use of the waveform interpolation involved in constructing surrogate models to perform the convolution analytically.

Chirp waveforms have been considered as toy examples to demonstrate the methodology. We have calculated the overlap between, and corresponding likelihood of, model templates and injected data to illustrate the smoothing properties of the technique. We also considered an ensemble of short MCMCs in the presence of white noise, using unsmooothed and smoothed waveform models: the smoothing process accelerates the convergence of the algorithm without increasing computational time.

In practice, this approach could be implemented in a similar way to simulated annealing: starting the MCMC with a large smoothing width and gradually reducing this to zero according to some predetermined schedule. Alternatively, many different chains could be run with different smoothing widths, along with interchain communication, analagous to parallel tempering. A comparison of these different methods, using higher-dimensional waveform models, will be investigated in future work.

\begin{acknowledgments}
RHC is supported by STFC. JG is supported by the Royal Society.
\end{acknowledgments}

\bibliography{likelihood-transform}

\appendix
\section{\label{app:PNphase}Binary phase to 3.5PN order}
To complete the description of the GWs emitted from a circular black hole binary used in section \ref{sec:examples}, we need an expression for the phase and frequency as a function of time. Using the auxilliary time variable
\begin{equation}
\tau(t) \equiv \frac{\nu c^3}{5GM}\left(t_c-t\right),
\end{equation}
where $t_c$ is the coalescence time of the binary, the phase can be computed to 3.5PN order as~\cite{Blanchet2002,Blanchet2004}
\begin{widetext}\begin{align}\label{eq:PNphase}
\Phi = - \frac1{\nu} \Bigg\{& \tau^{5/8} + \left( \frac{3715}{8064} + \frac{55}{96} \nu \right) \tau^{3/8} - \frac{3}{4}\pi\tau^{1/4} + \left( \frac{9275495}{14450688} + \frac{284875}{258048} \nu + \frac{1855}{2048} \nu^2 \right)\tau^{1/8}\nonumber\\
 &+ \left( -\frac{38645}{172032} + \frac{65}{2048} \nu\right) \pi \ln \left(\frac{\tau}{\tau_0}\right) + \left(\frac{831032450749357}{57682522275840}-\frac{53}{40}\pi^2-\frac{107}{56}\gamma_\mathrm{E}+\frac{107}{448}\ln\left(\frac{\tau}{256}\right)\right.\nonumber\\
 &+ \left.\left[-\frac{126510089885}{4161798144}+\frac{2255}{2048}\pi^2\right]\nu+\frac{154565}{1835008}\nu^2-\frac{1179625}{1769472}\nu^3\right)\tau^{-1/8}\nonumber\\
 &+ \left(\frac{188516689}{173408256}+\frac{488825}{516096}\nu-\frac{141769}{516096}\nu^2\right)\pi \tau^{-1/4}\Bigg\}\;.
\end{align}\end{widetext}
The constant $\tau_0$ sets the initial conditions when the binary passes some fiducial frequency; we choose $\tau_0 = \tau(0)$. $\gamma_\mathrm{E}$ is the Euler-Mascheroni constant.

The orbital frequency of the binary $\Omega = \dot{\Phi}$ can be computed by differentiating \eqref{eq:PNphase} with respect to $t$.

\end{document}